\documentclass{article}

\usepackage[ruled,vlined]{algorithm2e}

\usepackage{arxiv}

\usepackage[utf8]{inputenc} % allow utf-8 input
\usepackage[T1]{fontenc}    % use 8-bit T1 fonts
\usepackage{hyperref}       % hyperlinks
\usepackage{url}            % simple URL typesetting
\usepackage{booktabs}       % professional-quality tables
\usepackage{amsfonts}       % blackboard math symbols
\usepackage{nicefrac}       % compact symbols for 1/2, etc.
\usepackage{microtype}      % microtypography
\usepackage{lipsum}
\usepackage{graphicx}
\graphicspath{ {./images/} }
\usepackage{amsmath}
\usepackage{amssymb}
\usepackage{amsthm}
\usepackage{natbib}
\usepackage{caption}
\usepackage{arydshln} % for \cdashline
\usepackage{multirow}
\usepackage{subcaption}

\theoremstyle{definition}

\theoremstyle{plain}

\usepackage{xcolor}

\title{Neural Network-based Partial-Linear Single-Index Models for Environmental Mixtures Analysis}

\author{
 Hyungrok Do, Yuyan Wang, Mengling Liu, and Myeonggyun Lee\textsuperscript{*} \\[1.1ex]
  Division of Biostatistics\\
  Department of Population Health\\
  NYU Grossman School of Medicine\\
  New York, NY, 10016 \\
  \texttt{\{firstname.lastname\}@nyulangone.org} \\
  \textsuperscript{*}Corresponding author\\
}

\begin{document}
\maketitle
\begin{abstract}
Evaluating the health effects of complex environmental mixtures remains a central challenge in environmental health research. Existing approaches vary in their flexibility, interpretability, scalability, and support for diverse outcome types, often limiting their utility in real-world applications. To address these limitations, we propose a neural network-based partial-linear single-index (NeuralPLSI) modeling framework that bridges semiparametric regression modeling interpretability with the expressive power of deep learning. The NeuralPLSI model constructs an interpretable exposure index via a learnable projection and models its relationship with the outcome through a flexible neural network. The framework accommodates continuous, binary, and time-to-event outcomes, and supports inference through a bootstrap-based procedure that yields confidence intervals for key model parameters. We evaluated NeuralPLSI through simulation studies under a range of scenarios and applied it to data from the National Health and Nutrition Examination Survey (NHANES) to demonstrate its practical utility. Together, our contributions establish NeuralPLSI as a scalable, interpretable, and versatile modeling tool for mixture analysis. To promote adoption and reproducibility, we release a user-friendly open-source software package that implements the proposed methodology and supports downstream visualization and inference (\texttt{https://github.com/hyungrok-do/NeuralPLSI}).
\end{abstract}

% keywords can be removed

\section{Background}

Humans encounter a wide array of chemical and nonchemical agents throughout their lifespans \cite{wild2005complementing, stafoggia2017statistical, frederiksen2014human, aylward2013evaluation, exley2015pilot}. Carefully designed scientific studies, together with advances in exposure science and related technologies, have facilitated the investigation of the health effects of these complex mixtures \cite{hamra2018environmental, kortenkamp2007ten, braun2016can, kelley2019early, sanders2015perinatal}. Key research questions in environmental mixture analyses include: (i) overall effect estimation, what is the total impact of the mixture on the health outcome; (ii) identification of toxic agents, which specific congeners or chemicals are associated with the outcome, and which are most influential; and (iii) interactions and nonlinear relationships: do interactions exist among exposures, and is the exposure–response relationship nonlinear?

To address key questions, recent methodological developments have led to a range of statistical methods capable of analyzing environmental mixtures, including Bayesian kernel machine regression (BKMR) \cite{bobb2015bayesian, bobb2018statistical}, weighted quantile sum (WQS) regression \cite{carrico2015characterization, gennings2013cohort}, quantile-based g-computation (q-gcomp) \cite{keil2020quantile}, and partial-linear single-index (PLSI) models \cite{wang2020family}. Briefly, WQS regression derives a one-dimensional weighted sum score of the exposures that has a linear relationship with a continuous health outcome, under the assumption that all exposure effects act in the same direction. Although WQS has been generalized to accommodate several types of outcomes \cite{renzetti2016gwqs} and is widely used in practice \cite{vuong2020prenatal, christensen2013multiple, colicino2020per, czarnota2015assessment}, the assumption of directional homogeneity can be restrictive in real-world settings. q-gcomp combines the inferential simplicity of WQS regression with the flexibility of g-computation, a causal inference technique \cite{robins1986new}. Unlike WQS, q-gcomp allows for exposure effects in different directions. While both WQS and q-gcomp are parametric approaches, BKMR is a Bayesian nonparametric method designed to handle complex, nonlinear relationships between exposure mixtures and outcomes. Despite its flexibility and powerful visualization tools \cite{bobb2018statistical, vuong2020prenatal, domingo2019association, tanner2020environmental}, BKMR is limited by computational scalability challenges, interpretability issues, and limited support of implementations for non-continuous outcomes such as binary and time-to-event data.

The PLSI model is a family of semiparametric models that constructs a single index of exposures as a linear combination and allows for flexible, nonparametric relationships with the outcome through an unspecified link function \cite{hardle1989investigating, hardle1993optimal, ichimura1993semiparametric, carrico2015characterization, yu2002penalized, liang2010estimation}. This modeling strategy has been increasingly developed to identify important covariates and characterize their joint effects on a variety of outcome types, especially in biomedical and environmental health studies \cite{huang2006polynomial, sun2008polynomial, wang2020family, liu2021organophosphate, jin2021partially, wang2023semiparametric, charifson2024evaluating, lee2024partiala, lee2024partialb}. The PLSI framework allows associations between exposures and outcomes to be either positive or negative providing interpretable quantification of the direction and relative importance of covariates, and models these effects flexibly through a nonparametric link function. Wang et al. \cite{wang2020family} proposed a unified PLSI modeling framework to assess potentially nonlinear joint effects of environmental exposures across diverse outcome types, including continuous, categorical, time-to-event, and longitudinal data. PLSI models have received growing attention in environmental mixture analyses \cite{liu2021organophosphate, liu2022determinants, cajachagua2025prenatal}, as well as in methodological advancements in biomedical research, including extensions for time-varying survival data \cite{lee2024partiala, lee2024partialb}, mean residual life models \cite{jin2021partially}, functional modeling \cite{li2025dynamic}, distributed lag quantile model \cite{wang2023semiparametric} and nested case-control studies \cite{shang2013partially}.

In 2017, the National Institute of Environmental Health Sciences (NIEHS) launched the Powering Research through Innovative Methods for Mixtures in Epidemiology (PRIME) initiative, which has spurred significant advances in methodological development \cite{joubert2022powering}. PRIME-supported projects have contributed to a range of innovative approaches, including BKMR-based causal mediation analysis \cite{devick2022bayesian} and multiple index modeling \cite{mcgee2023bayesian}, among others. Yet, to the best of our knowledge, methods for mixture analysis that adapt deep learning techniques, offering scalability to high-dimensional and large-scale datasets, the capacity to automatically learn informative representations from data, and enhanced interpretability, have been relatively underexplored. Furthermore, the dissemination and practical adoption of existing methods remain limited by the lack of robust, user-friendly, and well-maintained software tools that can meet the increasing demands of modern data complexity and evolving analytic frameworks.

In this study, we propose a neural network-based partial-linear single-index (NeuralPLSI) modeling framework to advance mixture analysis by bridging semiparametric interpretability with the representational power of deep learning. Our approach retains the advantages of the classical PLSI framework, such as constructing an interpretable exposure index and accommodating nonlinear exposure–response relationships, while leveraging neural network architectures to model complex associations and enhance scalability. In addition, we develop and publicly release a well-documented and extensible software package that facilitates implementation and encourages broader adoption by environmental health researchers. This contribution aims to support the next generation of mixture modeling, enabling robust, scalable, and interpretable analysis in both research and applied settings.

This paper is organized as follows. Section 2 introduces the proposed NeuralPLSI modeling framework. Section 3 presents extensive simulation studies evaluating the performance of the NeuralPLSI methods under various scenarios. Section 4 demonstrates the application of the proposed approach to data from the National Health and Nutrition Examination Survey (NHANES). Finally, Section 5 concludes with a discussion of the findings and outlines directions for future research.

\begin{figure}[t]
    \centering
    \includegraphics[width=0.8\linewidth]{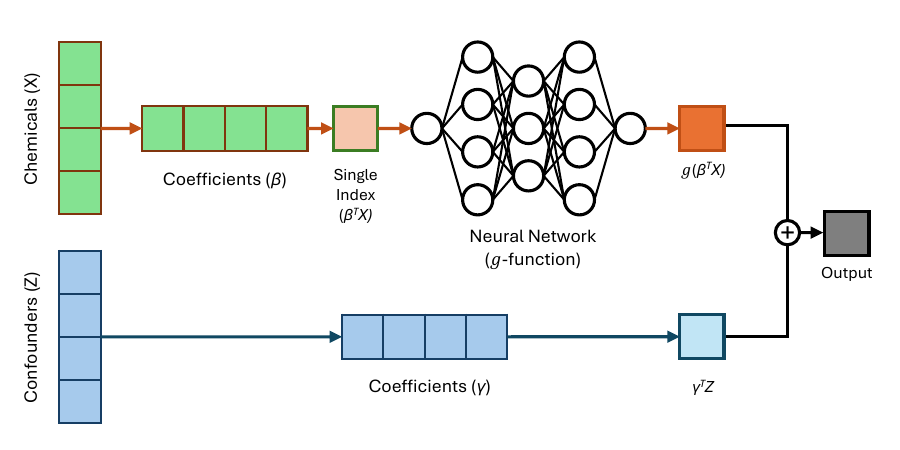}
    \caption{Overview of NeuralPLSI Model.}
    \label{fig:overview}
\end{figure}

\section{Methods}
\subsection{Neural network-based Partial-Linear Single-Index Model (NeuralPLSI)}
The partially linear single index (PLSI) model for continuous outcome is a semiparametric model of the form
\begin{equation}
    Y = g(\beta^{T}X) + \gamma^{T}Z + \varepsilon,
\end{equation}
where $Y \in \mathbb{R}$ denotes the continuous response variable, $X \in \mathbb{R}^{p}$ and $Z \in \mathbb{R}^{q}$ are covariate vectors, $\beta \in \mathbb{R}^{p}$ is the index direction, $\gamma \in \mathbb{R}^{q}$ is the vector of linear coefficients, $g:\mathbb{R} \to \mathbb{R}$ is an unknown linear or nonlinear function, and $\varepsilon$ is a zero-mean error term such that $\mathbb{E}[\varepsilon|X,Z] = 0$; extensions to other types of outcomes (e.g., binary and survival outcomes) are provided in Appendix A. The model assumes that the covariates in $Z$ contribute additively and linearly to the response, while the covariates in $X$ influence the outcome through a flexible univariate function of the linear projection $\beta^{T}X$. To ensure identifiability, it is common to impose constraints such as $\|\beta\|_{2} = 1$ and $\beta_{1} > 0$. This structure enables simultaneous modeling of linear and nonlinear effects and achieves dimension reduction by estimating the nonlinear component as a function of a one-dimensional index rather than the full $p$-dimensional vector $X$.
In this study, we propose to use a neural network $g_{\theta}:\mathbb{R} \to \mathbb{R}$ parameterized by $\theta$ to approximate the true nonlinear function $g$. We consider a learning process by solving the following optimization problem:
\begin{equation}
    \underset{\gamma, \beta, \theta}{\min} ~ \sum_{i=1}^{n} \Bigg(y_{i} - \Big(g_{\theta}(\beta^{T}x_{i}) + \gamma^{T}z_{i} \Big) \Bigg)^{2}.
\end{equation}
The problem minimizes the mean squared error on the training dataset $\{(x_{i}, z_{i}, y_{i}):i=1,\cdots,n\}$, with respect to the regression coefficient $\gamma$, index direction vector $\beta$, and neural network parameters $\theta$. As noted above, for the identifiability, we impose the constraint $\|\beta\|_{2} = 1$ and $\beta_{1} > 0$.
The use of a neural network to approximate the unknown function $g$ offers significant advantages in terms of flexibility and representational capacity. %According to the universal approximation theorem \cite{hornik1989multilayer,funahashi1989approximate}, a feedforward neural network with a finite number of neurons in a single hidden layer can approximate any Borel measurable function on a compact subset of $\mathbb{R}$, with arbitrary precision, provided that the activation function is non-constant, bounded, and continuous. This theoretical guarantee ensures that, with sufficient capacity and proper training, the neural network $g_{\theta}$ is capable of approximating the true nonlinear mapping $g(\beta^{T}X )$, regardless of its specific functional form. In contrast to traditional nonparametric methods such as kernel smoothing or spline basis expansions, which may struggle to adapt to complex local variations or require careful tuning of basis functions and bandwidth parameters, neural networks offer a data-driven, end-to-end solution that can flexibly learn intricate patterns and nonlinearities.
By classical universal approximation results \cite{hornik1989multilayer,funahashi1989approximate}, feedforward neural networks with nonpolynomial activation functions can approximate any continuous function on a compact subset of $\mathbb{R}$ to arbitrary accuracy. While this statement is used for theoretical motivation, in practice, $g_{\theta}$ may have an arbitary depth and structure (i.e., multiple layers). 
%By classical universal approximation results \cite{hornik1989multilayer,funahashi1989approximate}, a feedforward network with a single hidden layer and a nonpolynomial activation can approximate any continuous function on a compact subset of $\mathbb{R}$ to arbitrary accuracy. Consequently, with adequate capacity and training, $g_{\theta}$ can closely match the unknown mapping $g(\beta^{\top}X)$. 

Compared with kernel or spline methods, which often require careful bandwidth or knot selection and can be computationally demanding (for example, cross-validation over bandwidths and construction of $n\times n$ kernel matrices or large spline design matrices), neural networks provide a data driven end-to-end estimator that learns complex nonlinearities directly from the data. Stochastic gradient methods enable minibatch updates with per epoch cost proportional to $n$, avoid storing pairwise similarity matrices, and benefit from hardware acceleration. In practice, this yields near linear scaling in sample size and makes it feasible to estimate the index direction and the nonlinear component jointly in large-scale applications, while maintaining competitive approximation accuracy without extensive manual tuning (e.g., number of layers). 

When embedded within the PLSI framework, the neural network serves as a powerful and scalable estimator for the index-based nonlinear effect, enabling the model to accommodate highly nonlinear and non-smooth relationships between $ \beta^{T}X $ and $Y$, while still preserving the interpretability and structure afforded by the linear $\gamma^{T} Z $ term (Figure \ref{fig:overview}). To fix the additive nonidentifiability between the nonlinear and linear parts, we enforce a constraint $g(0)=0$ by augmenting the training objective with a small anchoring penalty. This centers the nonlinear effect at zero, prevents shifts from being absorbed into the intercept or $\gamma^{T}Z$, and preserves the interpretability of the linear component with negligible computational overhead.

To extend the scope of the NeuralPLSI model beyond continuous responses, we generalize its architecture to accommodate discrete and censored outcomes, such as binary, count, or time to event data (see Appendix A). This is achieved by incorporating a suitable link function, a foundational concept from the theory of Generalized Linear Models (GLMs) \cite{mccullagh2019generalized,nelder1972generalized}. The link function provides a monotonic transformation that maps the conditional mean of the response variable, $\mathbb{E}[Y | X, Z]$, to the model's additive predictor, $g_{\theta}(\beta^{T}X) + \gamma^{T}Z$. This allows the model, whose internal predictor is an unbounded real value, to align with the specific constraints of the outcome's distribution. For instance, mapping to the unit interval for a Bernoulli probability or to the positive real line for a Poisson rate. This approach thereby integrates the flexible representation capacity of neural networks with the established principles of statistical modeling, yielding a unified and versatile framework. We provide details of these extensions in Appendix A. 

Although the conceptual generalization is straightforward, the procedures for parameter estimation and statistical inference are nontrivial due to the model's nonconvex objective function and nonparametric components. Accordingly, a comprehensive treatment of the optimization algorithm and bootstrap inference methodology is provided in Appendix B.

\subsection{Simulation Settings}
We conducted simulations to evaluate the performance of the proposed NeuralPLSI method. We first generated eight exposures $X=(X_1, ..., X_8)^T$ from a multivariate normal distribution with zero mean vector and covariance matrix $\Sigma$ having diagonal entries as one and off-diagonal entries as $\rho=0.3$, assuming mild correlations between exposures. Continuous outcome $Y$ was generated from $Y=g(\beta^TX)+\gamma^TZ+\epsilon$, where $Z=(1, Z_1, Z_2, Z_3)^T$ with $Z_1$ and $Z_2$ following standard normal distribution and $Z_3$ following binomial distribution with $p=0.5$, $\epsilon \sim N(0,1)$. We set true parameters, $\beta=(1, 0.7, -0.5, 0.5, 0.3, -0.1, 0, 0)^T/\sqrt{(1^2+0.7^2+(-0.5)^2+0.3^2+(-0.1)^2)}$ and $\gamma=(1, 1, -0.5, 0.5)^T$. For the true link function $g$, we considered three scenarios: (1) linear, (2) S-shape, and (3) sigmoidal (see Supplementary Materials). In each setting, we generated 50 datasets, each consisting of $N=500$ and $2000$ observations. Using the generated datasets, we fitted the proposed NeuralPLSI and the classical PLSI model proposed by \cite{wang2020family}. 

We evaluated the performance of the model using bias, standard deviation (SD), standard error (SE) and coverage probability (CP) of the confidence interval (CI) 95\% for the parameters $\beta$ and $\gamma$. To compute standard errors, we used 100 bootstrapping samples. We also evaluated the estimated $g$ function with their 97.5\% and 2.5\% confidence bands in simulations using visualizations.

To evaluate the proposed NeuralPLSI models for other types of outcomes, we also generated binary and survival outcomes under the same settings described above. Details can be found in the supplementary materials, along with the simulation results.

\begin{table}[p]
\caption{Results of estimation metrics for coefficients using NeuralPLSI and PLSI methods across different true link function shapes for continuous outcome with $N=2000$. SD = Standard Deviation, SE = Standard Error, and CP = Coverage Probability.}\label{sim:results}

\centering
\footnotesize
\begin{tabular}{ccrrrrrrrr}
\toprule
\multirow{2}{*}{$g$\textbf{-function}} & \multirow{2}{*}{\textbf{Parameter}} & \multicolumn{4}{c@{\hskip 12pt}}{\textbf{NeuralPLSI}} & \multicolumn{4}{c}{\textbf{PLSI}} \\
\cmidrule(lr){3-6} \cmidrule(lr){7-10}
& & \multicolumn{1}{c}{\textbf{Bias}} & \multicolumn{1}{c}{\textbf{SD}} & \multicolumn{1}{c}{\textbf{SE}} & \multicolumn{1}{c}{\textbf{CP}} & \multicolumn{1}{c}{\textbf{Bias}} & \multicolumn{1}{c}{\textbf{SD}} & \multicolumn{1}{c}{\textbf{SE}} & \multicolumn{1}{c}{\textbf{CP}} \\
\midrule

\multirow{11}{*}{Linear} 
& $\beta_1$  & -0.0047 & 0.0240 & 0.0220 & 0.9600 & -0.0018 & 0.0157 & 0.0248 & 0.9800 \\
& $\beta_2$  & -0.0018 & 0.0264 & 0.0265 & 0.9800 & -0.0020 & 0.0236 & 0.0260 & 0.9600 \\
& $\beta_3$  & -0.0013 & 0.0257 & 0.0254 & 1.0000 &  0.0024 & 0.0213 & 0.0252 & 0.9400 \\
& $\beta_4$  & 0.0008 & 0.0326 & 0.0282 & 0.9200 &  0.0033 & 0.0256 & 0.0263 & 0.9400 \\
& $\beta_5$  & 0.0003 & 0.0298 & 0.0292 & 0.9800 & -0.0014 & 0.0239 & 0.0263 & 0.9600 \\
& $\beta_6$  & -0.0043 & 0.0258 & 0.0286 & 0.9600 &  0.0006 & 0.0233 & 0.0278 & 0.9200 \\
& $\beta_7$  & -0.0015 & 0.0301 & 0.0288 & 0.9800 & -0.0011 & 0.0272 & 0.0234 & 0.9200 \\
& $\beta_8$  & -0.0013 & 0.0254 & 0.0291 & 0.9800 &  0.0046 & 0.0266 & 0.0226 & 0.9400 \\
& $\gamma_1$ & -0.0084 & 0.0247 & 0.0252 & 0.9400 &  0.0013 & 0.0198 & 0.0225 & 0.8800 \\
& $\gamma_2$ & 0.0010 & 0.0249 & 0.0252 & 0.9800 &  0.0025 & 0.0234 & 0.0223 & 0.9400 \\
& $\gamma_3$ & -0.0020 & 0.0280 & 0.0250 & 0.9400 & -0.0051 & 0.0226 & 0.0223 & 0.9000 \\
\cdashline{1-10}

\multirow{11}{*}{s-Shape} 
& $\beta_1$  & -0.0005 & 0.0101 & 0.0098 & 0.9400 & -0.0011 & 0.0067 & 0.0111 & 0.9600 \\
& $\beta_2$  & -0.0015 & 0.0111 & 0.0119 & 1.0000 & -0.0004 & 0.0102 & 0.0115 & 0.9400 \\
& $\beta_3$  & 0.0002 & 0.0101 & 0.0114 & 1.0000 &  0.0000 & 0.0092 & 0.0101 & 0.9600 \\
& $\beta_4$  & 0.0020 & 0.0131 & 0.0127 & 0.9600 &  0.0013 & 0.0118 & 0.0111 & 0.9000 \\
& $\beta_5$  & -0.0012 & 0.0143 & 0.0130 & 0.9600 &  0.0004 & 0.0106 & 0.0115 & 0.9800 \\
& $\beta_6$  & -0.0015 & 0.0107 & 0.0127 & 0.9600 &  0.0007 & 0.0112 & 0.0113 & 0.9200 \\
& $\beta_7$  & 0.0002 & 0.0143 & 0.0130 & 0.9800 & -0.0002 & 0.0118 & 0.0116 & 0.9400 \\
& $\beta_8$  & -0.0019 & 0.0124 & 0.0130 & 0.9800 &  0.0028 & 0.0112 & 0.0116 & 0.9600 \\
& $\gamma_1$ & -0.0081 & 0.0263 & 0.0255 & 0.9600 &  0.0012 & 0.0086 & 0.0101 & 0.8800 \\
& $\gamma_2$ & 0.0008 & 0.0247 & 0.0252 & 0.9800 &  0.0024 & 0.0103 & 0.0111 & 0.9200 \\
& $\gamma_3$ & -0.0018 & 0.0277 & 0.0251 & 0.9400 & -0.0052 & 0.0101 & 0.0116 & 0.9200 \\
\cdashline{1-10}

\multirow{11}{*}{Sigmoid} 
& $\beta_1$  & -0.0005 & 0.0142 & 0.0135 & 0.9600 & -0.0022 & 0.0097 & 0.0159 & 0.9600 \\
& $\beta_2$  & -0.0024 & 0.0154 & 0.0163 & 1.0000 & -0.0003 & 0.0151 & 0.0164 & 0.9600 \\
& $\beta_3$  & 0.0012 & 0.0141 & 0.0156 & 1.0000 &  0.0002 & 0.0136 & 0.0161 & 0.9600 \\
& $\beta_4$  & 0.0026 & 0.0180 & 0.0175 & 0.9400 &  0.0020 & 0.0168 & 0.0166 & 0.9000 \\
& $\beta_5$  & -0.0010 & 0.0199 & 0.0179 & 0.9800 &  0.0007 & 0.0156 & 0.0164 & 0.9600 \\
& $\beta_6$  & -0.0027 & 0.0152 & 0.0175 & 0.9600 &  0.0011 & 0.0162 & 0.0161 & 0.9400 \\
& $\beta_7$  & -0.0008 & 0.0189 & 0.0178 & 0.9800 & -0.0011 & 0.0167 & 0.0166 & 0.9400 \\
& $\beta_8$  & -0.0027 & 0.0169 & 0.0179 & 0.9800 &  0.0045 & 0.0169 & 0.0166 & 0.9400 \\
& $\gamma_1$ & -0.0109 & 0.0255 & 0.0254 & 0.9400 &  0.0009 & 0.0124 & 0.0148 & 0.8600 \\
& $\gamma_2$ & 0.0014 & 0.0245 & 0.0252 & 0.9800 &  0.0023 & 0.0148 & 0.0160 & 0.9200 \\
& $\gamma_3$ & -0.0023 & 0.0276 & 0.0250 & 0.9400 & -0.0054 & 0.0145 & 0.0163 & 0.9200 \\
\bottomrule
\end{tabular}
\end{table}

\begin{figure}[p]
    \centering
    \includegraphics[width=0.95\linewidth]{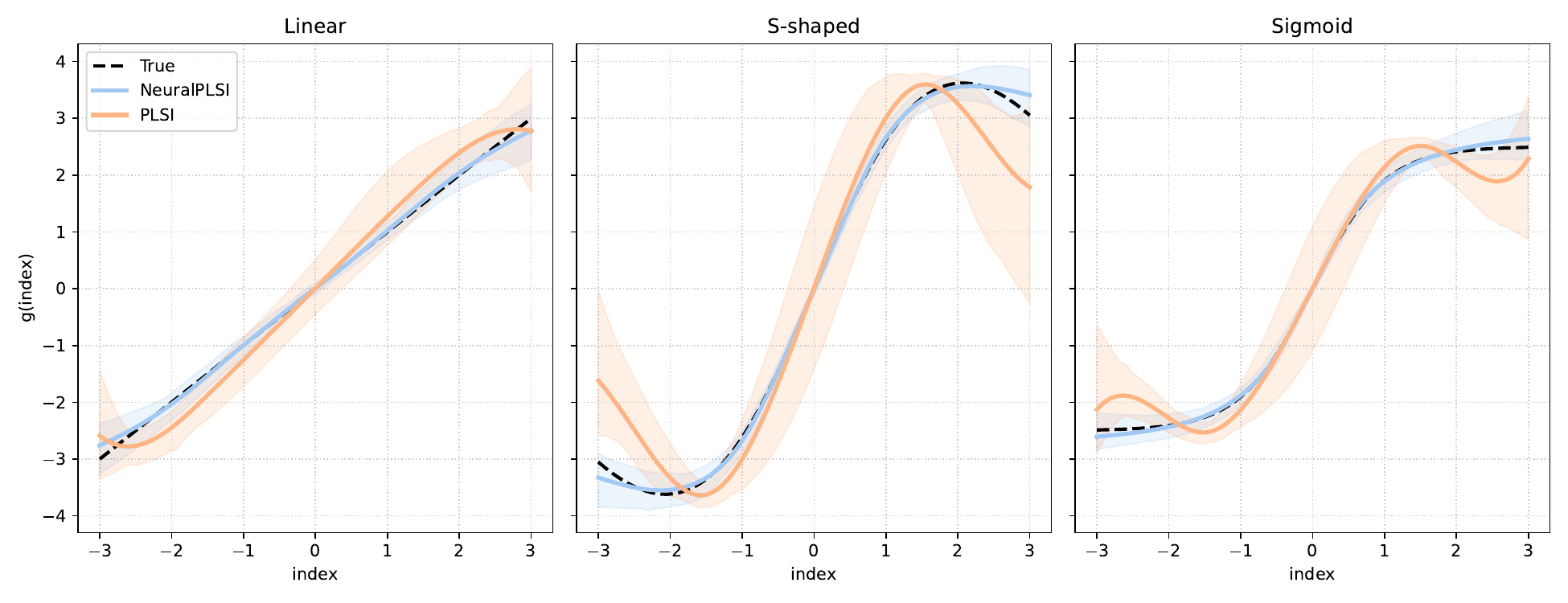}
    \caption{Estimated $g$ functions for the simulation study. }
    \label{fig:sim-cont-2000}
\end{figure}

\subsection{Modeling Serum Triglycerides with Environmental Mixtures in NHANES 2003-2004}
To demonstrate the proposed NeuralPLSI method for continuous outcome modeling in an environmental health context, we analyzed data from the 2003–2004 cycle of the National Health and Nutrition Examination Survey (NHANES) \cite{wang2020family}. NHANES is a nationally representative, cross-sectional survey designed to assess the health and nutritional status of the civilian, noninstitutionalized U.S. population through structured interviews, standardized physical examinations, and laboratory assessments. This survey includes extensive biomonitoring of environmental chemicals, making it a valuable resource for studying the health effects of chemical mixtures.

We followed the preprocessing pipeline described in \cite{wang2020family}, including participant inclusion criteria and exposure selection. The analytic sample consisted of 800 adults with complete data on serum triglyceride concentrations, eight selected environmental factors, and key demographic covariates. Inclusion required availability of laboratory measures for all selected exposures and triglycerides, as well as complete data on age, sex, and race/ethnicity. All participants provided written informed consent, and the study protocol was approved by the National Center for Health Statistics Institutional Review Board. The eight environmental exposures included $\alpha$-tocopherol, $\gamma$-tocopherol, retinyl palmitate, retinol, 3,3',4,4',5-pentachlorobiphenyl (3,3,4,4,5-pncb), PCB194, 2,3,4,6,7,8-hexachlorodibenzofuran (2,3,4,6,7,8-hxcdf), and trans-$\beta$-carotene. These analytes span lipid-soluble micronutrients such as tocopherols, retinoids, and carotenoids, as well as persistent organic pollutants (POPs) such as polychlorinated biphenyls and dioxin-like furans. Several of these chemicals are of public health interest because of their potential endocrine-disrupting properties, bioaccumulation, and long half-lives in human tissues. Tocopherols and carotenoids are antioxidants with possible cardiometabolic implications, while POPs have been linked to dyslipidemia, metabolic syndrome, and cardiovascular disease risk. To address right-skewed distributions, all exposure variables except retinol and the outcome were log-transformed, and all exposure variables standardized to have mean zero and unit variance. The covariates ($\mathbf{Z}$) included age in years, sex (male/female), and race/ethnicity (non-Hispanic White, non-Hispanic Black, Mexican American, other race including multiracial, and other Hispanic), all modeled linearly.

The health outcome, serum triglycerides (mg/dL), was modeled as a continuous variable without dichotomization. Elevated triglycerides are a well-established cardiovascular risk factor and an important marker of metabolic dysregulation \cite{miller2011triglycerides, boullart2012serum}. Environmental chemicals, particularly POPs, have been hypothesized to influence lipid metabolism through mechanisms such as oxidative stress, inflammation, and interference with nuclear receptor signaling \cite{sun2022review}.

In the NeuralPLSI framework, the eight exposures $\mathbf{X}$ were linearly combined into a single index to capture their joint effect on serum triglyceride levels, while the confounders $\mathbf{Z}$ entered the model linearly. The index was passed through a learned, fully nonparametric link function $g(\cdot)$, allowing flexible modeling of nonlinearities and interactions among exposures. This formulation retains interpretability through the index coefficients $\beta_j$, which describe the relative direction and magnitude of each exposure’s contribution, while extending beyond the constraints of traditional parametric linear regression.

\section{Results}

\subsection{Simulation Results}
Table \ref{sim:results} presents the results of our simulations under different true link functions with $N=2000$ (see Supplemental Materials for simulation results with $N=500$). Both the NeuralPLSI and classical PLSI models performed similarly, demonstrating empirical unbiasedness with reasonable efficiency. Both SD and SE using bootstrap samples were comparable between two methods. The coverage probabilities were also close to the nominal 95\% level. These findings were consistent regardless of whether the true link function was linear or nonlinear. Although the standard deviation (SD) and standard error (SE) slightly increased with $N=500$, which is expected, the proposed NeuralPLSI method maintained consistent performance (see Supplementary Materials).

Figure 2 illustrates the estimated link functions using our proposed NeuralPLSI model and the B-spline approach described by \cite{wang2020family}, along with their corresponding 95\% confidence intervals. Even though both methods approximated the true unknown link functions well in both linear and nonlinear cases, a closer inspection of the plots reveals a key advantage of the NeuralPLSI model, particularly in capturing the tail behavior of the link functions. While the B-psline PLSI model performed well in the central range, it exhibits high variance and a less precise fit at the tails of the index values. In contrast, our proposed NeuralPLSI model demonstrated a more stable and accurate approximation in these tail regions, as evidenced by its narrower confidence intervals and closer alighnment with the true link function. This superior performance in the tails could be particularly important in applicaitons where extreme values of the predictor index are clinically or scientifically significant. 
%shows the estimated link functions using the proposed neural network-based method and the B-spline approach described by \cite{wang2020family}, along with the corresponding 95\% confidence intervals. Both methods performed similarly and approximated the unknown link functions well in both the linear and nonlinear cases. 
For binary and survival outcomes, we observed similar simulation results of our proposed NeuralPLSI models, demonstrating empirically unbiased and reasonably efficient parameter estimates, along with well-approximated link functions (see Supplementary Materials).

\subsection{NeuralPLSI Analysis of Triglycerdies: NHANES 2003–2004}

% \begin{table}[t!]
% \small
% \caption{Results of NeuralPLSI and spline-based PLSI linear regression estimates with 95\% confidence intervals (CIs) for chemical exposures in the NHANES 2003–2004 triglyceride analysis.}
% \label{table:nhanes}
% \centering
% \begin{tabular}{ccccc}
% \hline
% Exposures & \multicolumn{2}{c}{NeuralPLSI} & \multicolumn{2}{c}{PLSI (\cite{wang2020family})} \\
% \cline{2-3} \cline{4-5}
%  & Estimate & 95\% CI & Estimate & 95\% CI \\
% \hline
% $\alpha$-Tocopherol & 0.591 & (0.520, 0.736) & 0.612 & (0.517, 0.707) \\
% $\gamma$-Tocopherol & 0.386 & (0.320, 0.498) & 0.400 & (0.326, 0.475) \\
% Retinyl-palmitate & 0.319 & (0.159, 0.403) & 0.386 & (0.289, 0.484) \\
% Retinol & 0.177 & (0.092, 0.275) & 0.154 & (0.080, 0.228) \\
% 3,3,4,4,5-pncb & 0.151 & (0.037, 0.205) & 0.093 & (0.018, 0.168) \\
% PCB194 & -0.262 & (-0.342, -0.134) & -0.258 & (-0.377, -0.138) \\
% 2,3,4,6,7,8-hxcdf & -0.349 & (-0.394, -0.202) & -0.266 & (-0.345, -0.186) \\
% trans-$\beta$-carotene & -0.395 & (-0.459, -0.279) & -0.383 & (-0.456, -0.310) \\
% \hline
% \end{tabular}
% \caption*{Note: Age, sex, and race/ethnicity, as potential confounders, were included as linear covariates. NeuralPLSI CIs were calculated using 1{,}000 bootstrap replicates. PLSI results are from the continuous-outcome analysis reported in Wang et al. (2020) \cite{wang2020family}}
% \end{table}

\begin{table}[t!]
\small
\caption{Results of NeuralPLSI and spline-based PLSI linear regression estimates with 95\% confidence intervals (CIs) for chemical exposures in the NHANES 2003–2004 triglyceride analysis.}
\label{table:nhanes}
\centering
\begin{tabular}{llcccc}
\hline
\multicolumn{2}{c}{\multirow{2}{*}{Covariates}}  & \multicolumn{2}{c}{NeuralPLSI} & \multicolumn{2}{c}{PLSI (\cite{wang2020family})} \\
\cline{3-4} \cline{5-6}
&  & Estimate & 95\% CI & Estimate & 95\% CI \\
\hline
\multirow{8}{*}{Exposures} 
 & $\alpha$-Tocopherol & 0.526 & (0.481, 0.700) & 0.612 & (0.517, 0.707) \\
 & $\gamma$-Tocopherol & 0.298 & (0.255, 0.453) & 0.400 & (0.326, 0.475) \\
 & Retinyl-palmitate & 0.368 & (0.175, 0.433) & 0.386 & (0.289, 0.484) \\
 & Retinol & 0.220 & (0.103, 0.291) & 0.154 & (0.080, 0.228) \\
 & 3,3,4,4,5-pncb & 0.165 & (0.042, 0.219) & 0.093 & (0.018, 0.168) \\
 & PCB194 & -0.269 & (-0.381, -0.158) & -0.258 & (-0.377, -0.138) \\
 & 2,3,4,6,7,8-hxcdf & -0.379 & (-0.427, -0.238) & -0.266 & (-0.345, -0.186) \\
 & trans-$\beta$-carotene & -0.455 & (-0.501, -0.308) & -0.383 & (-0.456, -0.310) \\
\hline
\multirow{10}{*}{Confounders} 
 & Age & 0.006 & (0.002, 0.009) & 0.005 & (0.001, 0.010) \\
 & Sex & & & & \\
 & \quad Male & Ref & & Ref & \\
 & \quad Female & -0.035 & (-0.176, 0.041) & -0.076 & (-0.167, 0.016) \\
 & Race/Ethnicity & & & & \\
 & \quad Non-Hispanic White & Ref & & Ref & \\
 & \quad Non-Hispanic Black & -0.203 & (-0.319, -0.059) & -0.138 & (-0.264, -0.011) \\
 & \quad Mexican American & 0.171 & (0.039, 0.291) & 0.175 & (0.054, 0.297) \\
 & \quad Other Race & 0.003 & (-0.128, 0.384) & 0.409 & (0.142, 0.676) \\
 & \quad Other Hispanic & 0.157 & (0.007, 0.429) & 0.355 & (0.083, 0.627) \\
\hline
\end{tabular}
\caption*{Note: Age, sex, and race/ethnicity were included as linear covariates. NeuralPLSI CIs were calculated using 1{,}000 bootstrap replicates. PLSI results are from the continuous-outcome analysis reported by \cite{wang2020family}.}
\end{table}

\begin{figure}[t]
    \centering
    \includegraphics[width=0.5\linewidth]{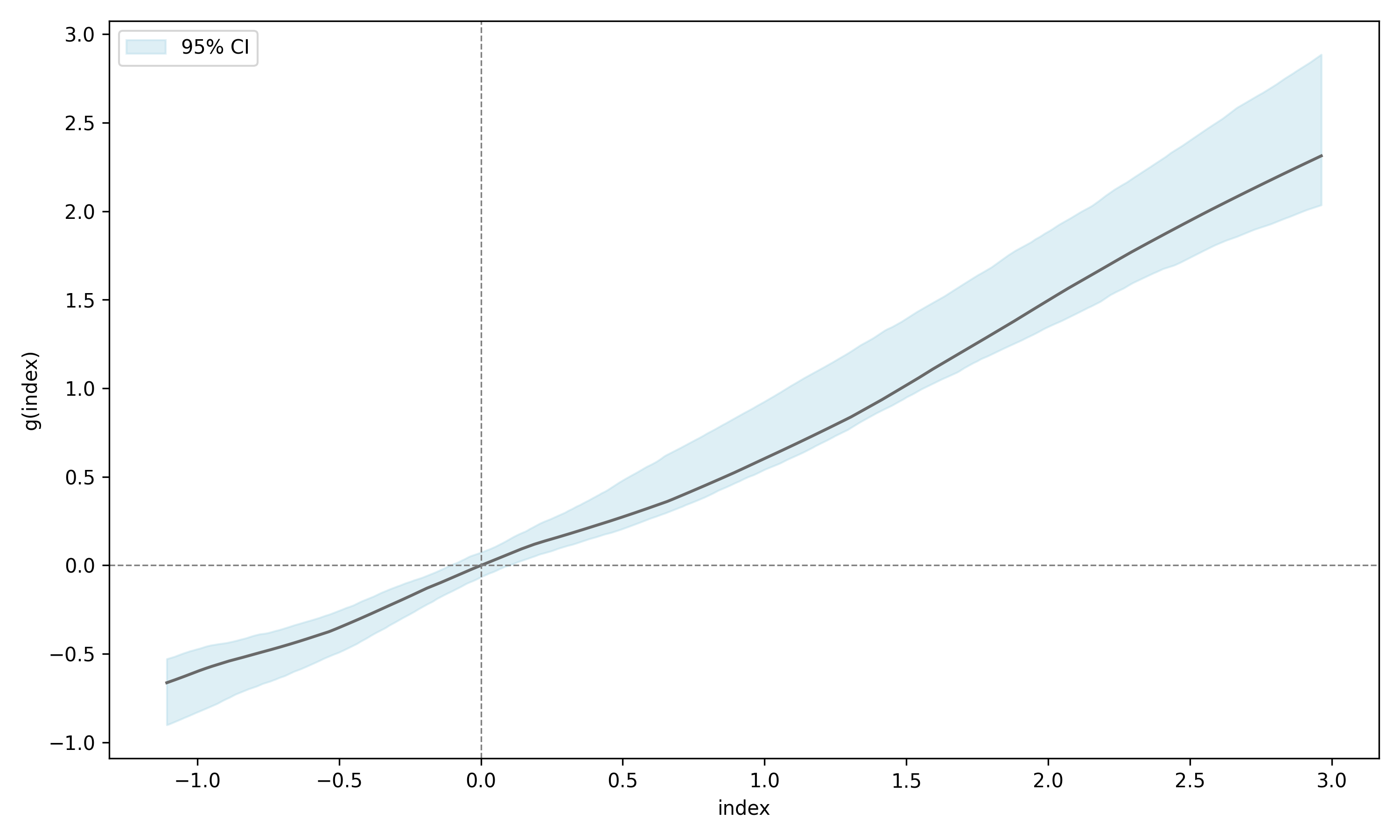}
    \caption{Estimated overall impact of 8 selected exposures on log-transformed triglycerides using the NeuralPLSI model.}
    \label{fig:tPLSI}
\end{figure}

For the continuous outcome of serum triglyceride concentration, the NeuralPLSI model identified both positive and negative associations between the eight environmental exposures and triglyceride levels (Table \ref{table:nhanes}). The estimated link function was monotone increasing (Figure \ref{fig:tPLSI}), indicating that the direction of each coefficient could be interpreted qualitatively in terms of its effect on mean triglyceride concentration.

%Among the exposures, $\alpha$-tocopherol showed the strongest positive association (estimate = 0.592, 95\% CI: 0.520 to 0.736), followed by $\gamma$-tocopherol (0.386, 95\% CI: 0.320 to 0.498), retinyl palmitate (0.319, 95\% CI: 0.159 to 0.403), retinol (0.177, 95\% CI: 0.092 to 0.275), and 3,3',4,4',5-pentachlorobiphenyl (0.151, 95\% CI: 0.037 to 0.205). In contrast, PCB194 (-0.262, 95\% CI: -0.342 to -0.134), 2,3,4,6,7,8-hexachlorodibenzofuran (-0.349, 95\% CI: -0.394 to -0.202), and trans-$\beta$-carotene (-0.395, 95\% CI: -0.459 to -0.279) were negatively associated with triglyceride concentrations. For the confounders, age showed a small but statistically significant positive association (0.232, 95\% CI: 0.088 to 0.299). Compared with non-Hispanic Whites, non-Hispanic Blacks (-0.168, 95\% CI: -0.318 to -0.047) had significantly lower triglyceride levels. Sex and other race/ethnicity contrasts showed no statistically significant associations in the NeuralPLSI analysis.

Among the exposures, $\alpha$-tocopherol showed the strongest positive association (estimate = 0.526, 95\% CI: 0.481 to 0.700), followed by retinyl palmitate (0.368, 95\% CI: 0.175 to 0.433), $\gamma$-tocopherol (0.298, 95\% CI: 0.255 to 0.453), retinol (0.220, 95\% CI: 0.103 to 0.291), and 3,3',4,4',5-pentachlorobiphenyl (0.165, 95\% CI: 0.042 to 0.219). In contrast, PCB194 (-0.269, 95\% CI: -0.381 to -0.158), 2,3,4,6,7,8-hexachlorodibenzofuran (-0.379, 95\% CI: -0.427 to -0.238), and trans-$\beta$-carotene (-0.455, 95\% CI: -0.501 to -0.308) were negatively associated with triglyceride concentrations. For the confounders, age showed a small but statistically significant positive association (0.006, 95\% CI: 0.002 to 0.009). Compared with non-Hispanic Whites, non-Hispanic Blacks (-0.203, 95\% CI: -0.319 to -0.059) had significantly lower triglyceride levels. Additionally, Mexican Americans (0.171, 95\% CI: 0.039 to 0.291) and Other Hispanics (0.157, 95\% CI: 0.007 to 0.429) showed significant positive associations, while sex and Other Race contrasts showed no statistically significant associations in the NeuralPLSI analysis.

When compared with the results of spline-based PLSI on NHANES 2003–2004, following \cite{wang2020family}, the overall pattern of associations was consistent in both direction and relative magnitude (Table \ref{table:nhanes}). In both analyses, $\alpha$-tocopherol emerged as the strongest positive contributor to the index, while trans-$\beta$-carotene was the strongest negative contributor. The sign patterns for all eight exposures matched those reported by \cite{wang2020family}, suggesting robust directional associations regardless of whether a spline-based or neural network-based link function was used. Differences were observed primarily in coefficient magnitudes: the NeuralPLSI estimates tended to be slightly larger in absolute value for the top-ranking exposures ($\alpha$- and $\gamma$-tocopherol, trans-$\beta$-carotene), possibly reflecting the model’s greater flexibility in capturing nonlinearities and complex interactions in the link function. The rank order of exposures by absolute coefficient size was identical between the two approaches, reinforcing the conclusion that the key drivers of triglyceride variability in this chemical mixture are largely consistent across modeling frameworks.

From an environmental health perspective, the positive associations for lipid-soluble antioxidants such as tocopherols and retinoids likely reflect their transport with triglyceride-rich lipoproteins. In contrast, the negative associations for POPs like PCB194, dioxin-like furans, and carotenoids may indicate more complex metabolic relationships or differential partitioning in lipid metabolism. The consistency between NeuralPLSI and spline-based PLSI results supports the robustness of these mixture–outcome relationships.

\section{Discussion}
We have developed a NeuralPLSI framework that unifies mixture analysis across common outcome types by combining an interpretable single exposure index with a learned nonlinear link. The approach preserves the dimensionality reduction and coefficient interpretability of classical single index models while adding flexibility through a neural link and scalability through stochastic optimization. A single training and inference pipeline adapts to continuous, categorical, and survival outcomes with modest changes that are limited to other mixture methods such as BKMR and WQS regression.

In our simulation study, NeuralPLSI provided a closer approximation to the true link functions than spline-based PLSI, particularly for nonlinear shapes such as S-shaped and sigmoid links at the extreme ends of the data. This demonstrates the advantage of a flexible neural network parameterization in capturing complex nonlinearities. However, the estimation of the index coefficients $\beta$ showed slightly higher variability (i.e., larger standard errors) with NeuralPLSI compared to spline-based PLSI. This likely arises because NeuralPLSI jointly estimates a highly flexible link function and the index coefficients, which increases the uncertainty in $\beta$ estimates. In contrast, spline-based PLSI uses a smoother, lower-dimensional representation of the link function, stabilizing $\beta$ estimation at the expense of potentially less accurate function approximation. Incorporating the covariance structure of exposures, with or without penalization, could improve the efficiency of $\beta$ estimates in the NeuralPLSI model, which we plan to explore in future work.

Methodologically, NeuralPLSI bridges the gap between predefined weighted sum approaches and fully nonparametric multivariate regressions. The single index provides a parsimonious and interpretable summary of the direction and relative magnitude of each exposure, while the learned link accommodates nonlinearity and interactions without manual basis design or pre-specification of interaction order. Compared with classical PLSI, NeuralPLSI reduces the tuning burden and scales naturally through minibatch optimization and automatic differentiation, which is particularly advantageous for larger cohorts and higher-dimensional exposure panels. Identifiability is enforced by constraining the index vector to unit norm with a fixed sign, and training is stabilized through standard regularization and early stopping. From a practical standpoint, the same objective function can incorporate case weights, facilitating the analysis of complex survey data when weighting is required. Post hoc visualizations of the fitted index response curve, together with confidence bands obtained via the bootstrap, further aid interpretation and model checking.

In the NHANES 2003–2004 case study of serum triglycerides modeled as a continuous outcome, NeuralPLSI identified mixed directional associations among the eight exposures. The strongest positive contribution was attributed to $\alpha$-tocopherol and the strongest negative contribution to trans $\beta$-carotene, consistent with established biological relationships. The estimated link was close to linear over the observed index range, which explains the similarity between NeuralPLSI and its parametric counterpart in this application while preserving the capacity to deviate from linearity when supported by the data.

This study has several limitations and potential avenues for extension. Although the single index reduces dimensionality, learning a flexible link still requires a sufficient sample size and careful regularization. The estimated link trades some immediate transparency for adaptability; index response plots, shape constraints when scientifically justified, and sensitivity analyses can help mitigate this concern. Sparsity-inducing or structure-aware penalties on the index can improve performance with higher-dimensional exposure panels or known chemical groupings. Multiple index generalizations can capture distinct mechanistic pathways acting through different linear combinations of exposures. The common training and bootstrap inference pipeline supports reproducible practice and allows extensions to higher dimensions and multiple indices. For survival analysis, allowing time-varying links or interactions with time can relax the proportional hazards assumption while preserving the interpretability of the exposure index. Promising future directions also include work on transportability across cohorts, subgroup performance assessment, shape-constrained links informed by toxicology, and NeuralPLSI-based causal mediation analysis.

%In summary, NeuralPLSI provides an interpretable and scalable approach for mixture analysis. Simulations show strong operating characteristics for continuous, binary, and survival outcomes, and the NHANES analysis demonstrates robust performance for continuous triglycerides with results consistent with prior biological knowledge. The common training and bootstrap inference pipeline supports reproducible practice and leaves room for extensions to higher dimensions and multiple indices.

\section{Conclusion}
Evaluating the health effects of complex environmental mixtures is a persistent challenge in epidemiologic research. We developed the NeuralPLSI modeling framework to combine semiparametric interpretability with the flexibility of deep learning, enabling construction of an interpretable exposure index and modeling of its relationship with diverse outcome types. Simulation studies and application to NHANES data demonstrated its scalability, interpretability, and practical utility for mixture analyses. The accompanying open source software package facilitates adoption, reproducibility, and broader application in environmental health research.

\section*{Availability of data and materials}
The data that support the findings in this paper are available from our previous study \cite{wang2020family}. A user-friendly open-source software package for the proposed NeuralPLSI models is available at (\texttt{https://github.com/hyungrok-do/NeuralPLSI}). 

\section*{Acknowledgement}
This work was partially supported by NIH/NIEHS grant R01ES032808 and CDC grants U01OH012637 and R21OH012793.

\bibliographystyle{unsrt}
\bibliography{references}

\newpage

%\section*{Acknowledgements}

% \section*{Author information}
% \subsection*{Authors and Affiliations}
% \textbf{Division of Biostatistics, Department of Population Health, New York University Grossman School of Medicine}
% Hyungrok Do, Yuyan Wang, Mengling Liu \& Myeonggyun Lee

% \subsection*{Contributions}
% H. Do, Y. Wang, M. Liu and M. Lee contributed to the conception and design of the study. H. Do and M. Lee performed the simulation study and data analysis. All authors participated in data interpretation, writing, and revision of the manuscript and approved the final version of this manuscript. 

% \section*{Ethics declarations}
% \subsection*{Ethics approval and consent to participate}
% Not applicable.

% \subsection*{Consent for publication}
% Not applicable. 

% \subsection*{Competing interests}
% The authors declare no competing interests. 

\section*{Appendix}
\subsection*{Appendix A. Extension of NeuralPLSI to Non-continuous Outcomes}
The NeuralPLSI model can be naturally extended beyond continuous outcomes by introducing a suitable link function that connects the expected response to the additive model components. This generalization places NeuralPLSI within the broader family of generalized neural network-based partial-linear single-index models, allowing its application to binary, count, and time-to-event outcomes.

In generalized linear models (GLMs), the conditional mean of the response variable $Y$ given covariates $(X, Z)$ is linked to a linear predictor through a monotonic, differentiable link function $\psi(\cdot)$. We extend this principle to our framework by modeling
\begin{equation}
    \mathbb{E}[Y | X, Z] = \psi^{-1}\left(g_{\theta}(\beta^{T}X) + \gamma^{T}Z\right),
\end{equation}
where $\psi^{-1}(\cdot)$ is the inverse link function appropriate for the distribution of $Y$. This formulation allows us to retain the flexible, nonparametric modeling of the index component $g(\beta^{T}X)$, while incorporating outcome-specific characteristics through $\psi$.

For example, when $Y \in \{0,1\}$, a common choice is the logistic link $\psi(\mu) = \log\left(\mu / (1 - \mu)\right)$, leading to a semiparametric single-index logistic regression:
\begin{equation}
    \text{logit}(\mathbb{P}(Y = 1 | X,Z)) = g_{\theta}(\beta^{T}X) + \gamma^{T}Z.
\end{equation}

For Poisson-distributed $Y \in \{0,1,2,\ldots\}$, we use a log link $\psi(\mu) = \log \mu$, yielding
\begin{equation}
    \log \mathbb{E}[Y | X,Z] = g_{\theta}(\beta^{T}X) + \gamma^{T}Z.
\end{equation}

The optimization problem in this setting becomes the maximization of a generalized log-likelihood:
\begin{equation}
    \underset{\gamma, \beta, \theta}{\max} ~ \sum_{i=1}^{n} \log p\left(y_{i} | \psi^{-1}\left(g_{\theta}(\beta^{T}x_{i}) + \gamma^{T}z_{i} \right) \right),
\end{equation}
where $p(y | \mu)$ is the conditional density or probability mass function of $Y$ under its assumed distribution (e.g., Bernoulli or Poisson).

The NeuralPLSI framework can also be extended to time-to-event data using the Cox proportional hazards model \cite{cox1972regression}. The survival variant fit naturally into the framework by embedding the single index within a proportional hazards formulation, allowing partial likelihood training with the same identifiability constraints and regularization. In this context, we model the hazard function $\lambda(t | X,Z)$ as
\begin{equation}
    \lambda(t | X, Z) = \lambda_0(t) \exp\left(g_{\theta}(\beta^{T}X) + \gamma^{T}Z\right),
\end{equation}
where $\lambda_0(t)$ is the baseline hazard function and the index-based nonlinear transformation $g_{\theta}(\beta^{T}X)$ replaces the usual linear predictor in standard Cox models. The model retains the proportional hazards structure while permitting nonlinear interactions among $X$ through the learned index.

Given a dataset $\{(x_i, z_i, o_i, \delta_i)\}$ with covariates $(x_i, z_i)$, observed times $o_i$, and event indicators $\delta_i$, we maximize the partial log-likelihood as:
\begin{equation}
    \underset{\gamma, \beta, \theta}{\max} \quad \sum_{i:\delta_i = 1} \left[ g_{\theta}(\beta^{T}x_{i}) + \gamma^{T}z_{i} - \log \sum_{j \in \mathcal{R}(t_i)} \exp\left(g_{\theta}(\beta^{T}x_{j}) + \gamma^{T}z_{j} \right) \right],
\end{equation}
where $\mathcal{R}(t_i)$ denotes the risk set at time $t_i$. This model enables flexible modeling of survival data with both interpretable linear effects and complex nonlinear components. We further note that this can easily be extended to the accelerated failure time framework or nonparametric survival models.

\subsection*{Appendix B. Optimization and Inference}

\begin{algorithm}[t]
\caption{Training Procedure for NeuralPLSI}
\KwIn{Training data $\{(x_i, z_i, y_i)\}_{i=1}^n$, learning rate $\eta$, number of epochs $E$, batch size $B$}
\KwOut{Trained parameters: $\gamma$, $\beta$, $\theta$}
Initialize $\gamma$, $\beta$ with $\|\beta\|_2 = 1$, and $\theta$\;
\For{epoch = 1 \KwTo $E$}{
    Shuffle the training data\;
    \For{each minibatch $\mathcal{B} = \{(x_b, z_b, y_b)\}_{b=1}^B$}{
        Compute $\hat{y}_b = g_\theta(\beta^T x_b) + \gamma^T z_b$\;
        Compute loss: $\mathcal{L}_{\mathcal{B}} = -\frac{1}{B} \sum_b \ell(y_b, \hat{y}_b) + \frac{1}{B} \sum_{b}(\beta^{T}x_{b})^{2}$\;
        Compute gradients $\nabla_\gamma \mathcal{L}_{\mathcal{B}}, \nabla_\beta \mathcal{L}_{\mathcal{B}}, \nabla_\theta \mathcal{L}_{\mathcal{B}}$\;
        Update parameters: $\gamma \leftarrow \gamma - \eta \nabla_\gamma$, $\beta \leftarrow \beta - \eta \nabla_\beta$, $\theta \leftarrow \theta - \eta \nabla_\theta$\;
        \If{$\beta_{1} < 0$}{
        Flip the sign of $\beta$: $\beta \gets -\beta $\;
        (Optional) Flip the sign of momentum or exponential average w.r.t $\beta$\;
        }
        Project $\beta \leftarrow \beta / \|\beta\|_2$\;
    }
}
\Return $\gamma$, $\beta$, $\theta$\;
\end{algorithm}

Because the objective function of NeuralPLSI involves a composition of a nonlinear neural network $g_\theta(\beta^{T} x)$ with parameters $\theta$ and a learnable projection vector $\beta$, the resulting optimization problem is highly non-convex. This non-convexity stems from multiple sources. First, the neural network $g_\theta(\cdot)$ introduces complex, non-linear mappings that can lead to a highly non-convex loss surface with many local minima and saddle points \cite{choromanska2015loss, dauphin2014identifying}. Second, the interaction between the neural network and the index direction $\beta$ compounds this complexity. Since the output of the neural network depends on the projection $\beta^{T} x$, the loss is sensitive not only to the orientation of $\beta$ but also to the internal structure of $g_\theta$, creating a coupled optimization landscape.

To solve this type of problem, stochastic gradient descent (SGD) and its variants, such as Adam \citep{kingma2014adam}, are popularly used. These methods are widely recognized as effective for optimizing deep neural networks with highly non-convex objectives. Unlike full-batch gradient descent, which computes gradients over the entire dataset at each iteration, SGD uses randomly sampled minibatches to estimate the gradient direction. This stochasticity serves two crucial purposes. First, it drastically reduces memory consumption, making training feasible in large-scale datasets and resource-constrained environments. Second, and more importantly, in the context of the NeuralPLSI, the inherent noise introduced by minibatch updates acts as a regularizer that helps the optimizer escape poor local minima and saddle points. In high-dimensional non-convex landscapes, saddle points, where gradients vanish but the point is not a local minimum, are far more prevalent than local minima and often pose significant obstacles to optimization \citep{dauphin2014identifying}. Full-batch gradient descent can become stuck near such saddle points because of vanishing gradients and the curvature being flat in some directions and negative in others. SGD, by contrast, introduces sufficient perturbations through minibatch noise to break symmetry and escape these degenerate regions, allowing continued progress toward lower loss values. Moreover, non-convex loss surfaces often contain multiple local minima, some of which are sharp and narrow, while others are flat and wide. Sharp minima correspond to solutions with high curvature, where small changes in parameters can lead to large changes in loss, often resulting in poor generalization. Flat minima, on the other hand, tend to be more robust and yield better generalization performance \citep{hochreiter1997flat, keskar2017large}. The stochasticity in SGD biases the search trajectory toward flatter regions of the loss surface, improving the generalizability and stability of the model \citep{jastrzebski2017three, neyshabur2017exploring}. These characteristics make stochastic optimization particularly well-suited for NeuralPLSI, where the composition of a nonlinear neural network and a learnable projection direction creates a highly coupled and rugged loss surface. The ability of SGD and its adaptive variants to escape saddle points, avoid sharp minima, and promote generalizable solutions is essential for reliable training in such settings.

Training of neural networks is sensitive to initialization, and poor initialization can lead to slow convergence or suboptimal solutions. We use standard initialization schemes such as He initialization \cite{he2015delving} to ensure that the variance of activations is preserved across layers, thereby stabilizing gradient propagation during early training epochs.

During training, the model parameters $(\beta, \gamma, \theta)$ are jointly updated via gradient-based optimization to minimize a loss function, typically mean squared error or negative log-likelihood, depending on the outcome type. After each gradient update, we re-normalize the vector $\beta$ to enforce the constraint $\|\beta\|_2 = 1$ and flip the signs of $\beta$ if $\beta_1 < 0$. This constraint is essential for identifiability: without it, arbitrary rescaling between $\beta$ and the input to $g_\theta$ could yield the same overall model behavior but with different parameter values.

In summary, the training of NeuralPLSI models integrates (i) stochastic minibatch updates for memory-efficient and noise-assisted optimization, (ii) adaptive optimizers for scale-aware learning, (iii) careful weight initialization to preserve stability, and (iv) constraint enforcement for identifiability. This combination allows effective learning even in complex and high-dimensional data regimes.

Conventional statistical models such as linear regression, GLMs, and Cox proportional hazards models permit formal statistical inference on model parameters through asymptotic theory. However, in the NeuralPLSI framework, inference is considerably more challenging due to the nonparametric nature of the neural network component and the absence of closed-form MLEs.

The NeuralPLSI model has the general form $f(x) = g(\beta^{T} x) + \gamma^{T} z$ where $\beta \in \mathbb{R}^p$ and $\gamma \in \mathbb{R}^q$ are learnable linear parameters, and $g: \mathbb{R} \to \mathbb{R}$ is a nonlinear function modeled via a neural network. Despite extensive progress in quantifying predictive uncertainty in neural networks, such as MC dropout, SWAG, and Laplace approximations, these approaches do not directly yield standard errors or confidence intervals for $\beta$ and $\gamma$. Analytical variance estimation is further complicated by the lack of a well-specified likelihood and the interdependence of $\beta$, $\gamma$, and $g(\cdot)$ in the optimization process.

Therefore, to enable inference of the parametric components $\beta$ and $\gamma$, we adopt a nonparametric bootstrap procedure (Algorithm \ref{pseudo:bootstrap}). This approach is model-agnostic and accommodates the complexities inherent to neural network optimization. Let $\hat{\beta}$ and $\hat{\gamma}$ denote the parameter estimates obtained by fitting the NeuralPLSI model to the complete dataset $\mathcal{D}_n = \{(x_i, y_i)\}_{i=1}^n$.

\begin{algorithm}[t]
    \caption{Bootstrap Inference for NeuralPLSI}
    \label{pseudo:bootstrap}
    \KwIn{Original dataset $\mathcal{D}_n = \{(x_i, y_i)\}_{i=1}^n$, number of bootstrap replicates $B$}
    \KwOut{Bootstrap estimates of $\{\hat{\beta}^{(b)}, \hat{\gamma}^{(b)}\}_{b=1}^B$, standard errors, confidence intervals}
    \For{$b = 1$ \KwTo $B$}{
        Sample with replacement $n$ pairs from $\mathcal{D}_n$ to construct bootstrap dataset $\mathcal{D}_n^{(b)}$\;
        Fit the NeuralPLSI model on $\mathcal{D}_n^{(b)}$ to obtain estimates $\hat{\beta}^{(b)}$, $\hat{\gamma}^{(b)}$\;
    }
    Compute bootstrap means: $\bar{\beta} = \frac{1}{B} \sum_{b=1}^B \hat{\beta}^{(b)}$, $\bar{\gamma} = \frac{1}{B} \sum_{b=1}^B \hat{\gamma}^{(b)}$\;
    Compute standard errors: $\widehat{\mathrm{SE}}(\hat{\beta}_j) = \sqrt{\frac{1}{B-1} \sum_{b=1}^B (\hat{\beta}^{(b)}_j - \bar{\beta}_j)^2}$, and similarly for $\hat{\gamma}_j$\;
    Construct confidence intervals using either the percentile or the normal approximation method\;
    \Return $\{\hat{\beta}^{(b)}, \hat{\gamma}^{(b)}\}_{b=1}^B$, estimated standard errors, confidence intervals\;
\end{algorithm}

Confidence intervals for each component can be obtained via the bootstrap percentile method or by assuming approximate normality. For the $j$-th component of $\hat{\beta}$, the confidence interval is given by
\begin{equation}
    \hat{\beta}_j \pm z_{1 - \alpha/2} \cdot \widehat{\mathrm{SE}}(\hat{\beta}_j),
\end{equation}
and similarly, for the $k$-th component of $\hat{\gamma}$, the confidence interval is
\begin{equation}
    \hat{\gamma}_k \pm z_{1 - \alpha/2} \cdot \widehat{\mathrm{SE}}(\hat{\gamma}_k),
\end{equation}
where $z_{1 - \alpha/2}$ is the $(1 - \alpha/2)$-quantile of the standard normal distribution, and $\alpha$ denotes the desired significance level (e.g., $\alpha = 0.05$ for a $95\%$ confidence interval).

The above inference procedure generalizes naturally to variants of the NeuralPLSI model involving non-continuous responses, such as those modeled via GLM or Cox proportional hazards models. In such settings, the linear component $\gamma^T x$ and the nonlinear index $g(\beta^T x)$ are combined within appropriate link or hazard functions, and inference on $(\beta, \gamma)$ can still be validly conducted using nonparametric bootstrap methods \citep{tibshirani1993introduction,shao2012jackknife}. This bootstrap-based approach enables robust estimation of uncertainty in partially linear models involving neural networks. It is flexible, easy to implement, and does not require explicit modeling assumptions about the distribution of the estimators. Moreover, it provides valid inference for both linear and nonlinear components across continuous, discrete, and censored outcome settings.

\subsection*{Appendix C. True Link Function Settings for Simulation Study}
\begin{figure}[h]
    \centering
    \includegraphics[width=0.95\linewidth]{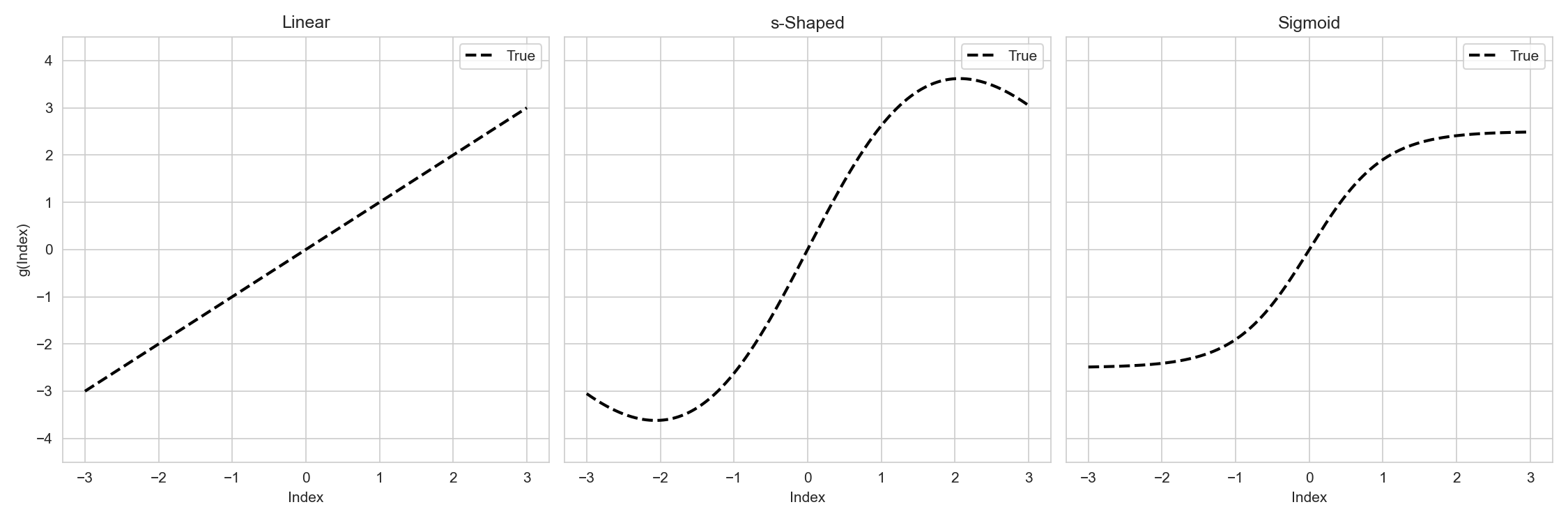}
    \caption{Shapes of true link functions in simulation study.}
    \label{figureS1}
\end{figure}

In our simulation study, we assumed one linear and two nonlinear link functions as follows. $g_1$ was assumed to be a S-shaped function, that is, $g_1(s)=10(\frac{2}{1+exp(-s)}-0.2s-1)$. $g_3$ was assumed as a sigmoidal function, $g_3(s)=5(\frac{1}{1+exp(-2s)}-0.5)$. These two shape functions were refereed from the simulation setting of \cite{mcgee2023bayesian}. We provided the shapes of each nonlinear function in Figure \ref{figureS1}. 

\subsection*{Appendix D. Additional Simulation Results}
Table 1 presents the simulation results for a continuous outcome with $N=500$. Both NeuralPLSI and PLSI models performed similarly, yielding empirically unbiased estimates and reasonable efficiency. Compared to the results for $N=2000$ (Table 1 and Figure 2 of the manuscript), SD and SE increased, resulting in slightly underperformed CP relative to the 95\% nominal level, as expected. Similar patterns were observed for the binary and survival outcomes (Tables 2–4 and Figures 3–5).

\begin{table}[p]
\caption{Results of estimation metrics for coefficients using NeuralPLSI method across different true link functions for continuous outcome with $N=500$ and $N=2000$. SD = Standard Deviation, SE = Standard Error, and CP = Coverage Probability.}
\label{sim:results-cont-combined}
\centering
\footnotesize
\begin{tabular}{ccrrrrrrrr}
\toprule
\multirow{2}{*}{\textbf{$g$-function}} & \multirow{2}{*}{\textbf{Parameter}} &
\multicolumn{4}{c@{\hskip 12pt}}{\textbf{NeuralPLSI ($N=500$)}} &
\multicolumn{4}{c}{\textbf{NeuralPLSI ($N=2000$)}} \\
\cmidrule(lr){3-6} \cmidrule(lr){7-10}
& & \textbf{Bias} & \textbf{SD} & \textbf{SE} & \textbf{CP} & \textbf{Bias} & \textbf{SD} & \textbf{SE} & \textbf{CP} \\
\midrule
\multirow{11}{*}{Linear} 
& $\beta_1$  & -0.0144 & 0.0475 & 0.0534 & 0.9800 & -0.0047 & 0.0240 & 0.0220 & 0.9600 \\
& $\beta_2$  & -0.0107 & 0.0507 & 0.0611 & 1.0000 & -0.0018 & 0.0264 & 0.0265 & 0.9800 \\
& $\beta_3$  & 0.0085 & 0.0694 & 0.0581 & 0.9400 & -0.0013 & 0.0257 & 0.0254 & 1.0000 \\
& $\beta_4$  & 0.0065 & 0.0558 & 0.0645 & 1.0000 & 0.0008 & 0.0326 & 0.0282 & 0.9200 \\
& $\beta_5$  & 0.0022 & 0.0824 & 0.0664 & 0.9800 & 0.0003 & 0.0298 & 0.0292 & 0.9800 \\
& $\beta_6$  & -0.0006 & 0.0615 & 0.0642 & 1.0000 & -0.0043 & 0.0258 & 0.0286 & 0.9600 \\
& $\beta_7$  & -0.0094 & 0.0618 & 0.0652 & 0.9600 & -0.0015 & 0.0301 & 0.0288 & 0.9800 \\
& $\beta_8$  & 0.0012 & 0.0604 & 0.0653 & 0.9800 & -0.0013 & 0.0254 & 0.0291 & 0.9800 \\
& $\gamma_1$  & -0.0939 & 0.0529 & 0.0997 & 0.8400 & -0.0084 & 0.0247 & 0.0252 & 0.9400 \\
& $\gamma_2$  & 0.0119 & 0.0398 & 0.0578 & 1.0000 & 0.0010 & 0.0249 & 0.0252 & 0.9800 \\
& $\gamma_3$  & -0.0073 & 0.0458 & 0.0571 & 0.9800 & -0.0020 & 0.0280 & 0.0250 & 0.9400 \\
\cdashline{1-10}
\multirow{11}{*}{s-Shape} 
& $\beta_1$  & -0.0037 & 0.0203 & 0.0248 & 0.9600 & -0.0005 & 0.0101 & 0.0098 & 0.9400 \\
& $\beta_2$  & -0.0023 & 0.0209 & 0.0343 & 1.0000 & -0.0015 & 0.0111 & 0.0119 & 1.0000 \\
& $\beta_3$  & 0.0007 & 0.0298 & 0.0293 & 0.9400 & 0.0002 & 0.0101 & 0.0114 & 1.0000 \\
& $\beta_4$  & 0.0032 & 0.0234 & 0.0336 & 0.9800 & 0.0020 & 0.0131 & 0.0127 & 0.9600 \\
& $\beta_5$  & 0.0004 & 0.0351 & 0.0318 & 0.9600 & -0.0012 & 0.0143 & 0.0130 & 0.9600 \\
& $\beta_6$  & 0.0020 & 0.0253 & 0.0307 & 1.0000 & -0.0015 & 0.0107 & 0.0127 & 0.9600 \\
& $\beta_7$  & -0.0044 & 0.0293 & 0.0300 & 0.9400 & 0.0002 & 0.0143 & 0.0130 & 0.9800 \\
& $\beta_8$  & 0.0025 & 0.0267 & 0.0304 & 0.9800 & -0.0019 & 0.0124 & 0.0130 & 0.9800 \\
& $\gamma_1$  & -0.0941 & 0.0492 & 0.0992 & 0.8800 & -0.0081 & 0.0263 & 0.0255 & 0.9600 \\
& $\gamma_2$  & 0.0138 & 0.0408 & 0.0584 & 1.0000 & 0.0008 & 0.0247 & 0.0252 & 0.9800 \\
& $\gamma_3$  & -0.0101 & 0.0453 & 0.0581 & 0.9800 & -0.0018 & 0.0277 & 0.0251 & 0.9400 \\
\cdashline{1-10}
\multirow{11}{*}{Sigmoid} 
& $\beta_1$  & -0.0057 & 0.0266 & 0.0317 & 0.9800 & -0.0005 & 0.0142 & 0.0135 & 0.9600 \\
& $\beta_2$  & -0.0032 & 0.0292 & 0.0394 & 1.0000 & -0.0024 & 0.0154 & 0.0163 & 1.0000 \\
& $\beta_3$  & 0.0042 & 0.0408 & 0.0362 & 0.9400 & 0.0012 & 0.0141 & 0.0156 & 1.0000 \\
& $\beta_4$  & 0.0037 & 0.0331 & 0.0409 & 0.9800 & 0.0026 & 0.0180 & 0.0175 & 0.9400 \\
& $\beta_5$  & 0.0012 & 0.0493 & 0.0414 & 0.9600 & -0.0010 & 0.0199 & 0.0179 & 0.9800 \\
& $\beta_6$  & 0.0029 & 0.0357 & 0.0400 & 1.0000 & -0.0027 & 0.0152 & 0.0175 & 0.9600 \\
& $\beta_7$  & -0.0062 & 0.0415 & 0.0397 & 0.9400 & -0.0008 & 0.0189 & 0.0178 & 0.9800 \\
& $\beta_8$  & 0.0023 & 0.0387 & 0.0402 & 0.9800 & -0.0027 & 0.0169 & 0.0179 & 0.9800 \\
& $\gamma_1$  & -0.1026 & 0.0501 & 0.1013 & 0.8600 & -0.0109 & 0.0255 & 0.0254 & 0.9400 \\
& $\gamma_2$  & 0.0146 & 0.0402 & 0.0576 & 1.0000 & 0.0014 & 0.0245 & 0.0252 & 0.9800 \\
& $\gamma_3$  & -0.0093 & 0.0445 & 0.0573 & 0.9800 & -0.0023 & 0.0276 & 0.0250 & 0.9400 \\
\bottomrule
\end{tabular}
\end{table}

\begin{figure}[p]
    \centering
    \includegraphics[width=0.95\linewidth]{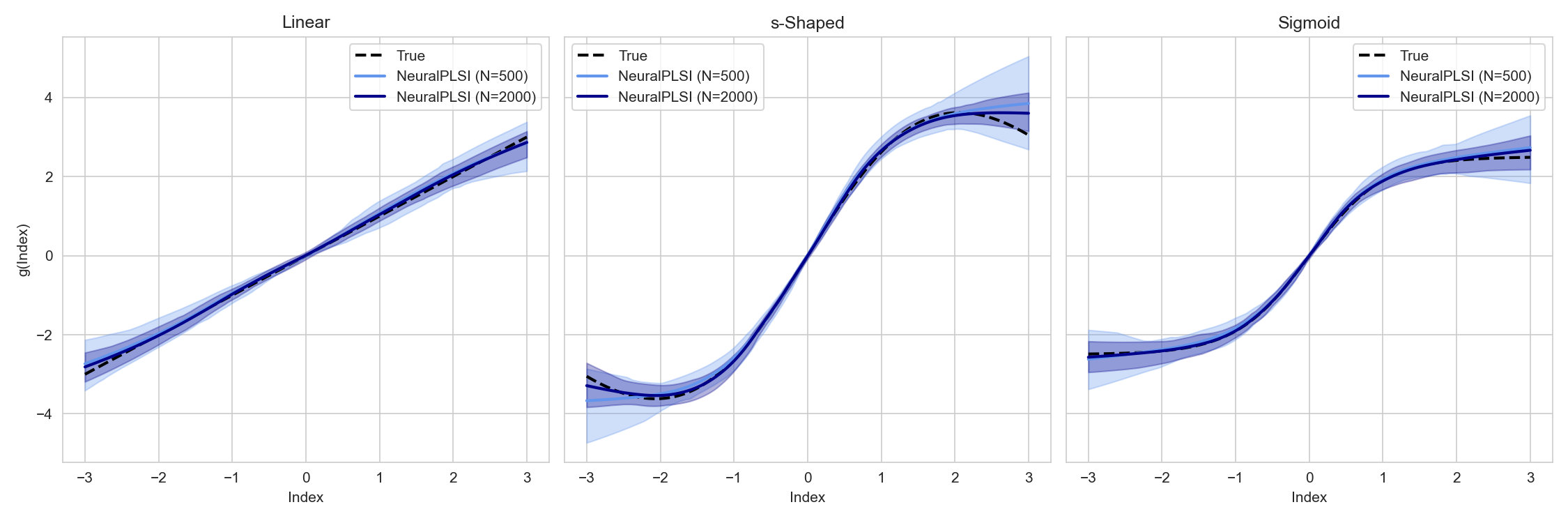}
    \caption{Estimated $g$ functions for the simulation study.}
    \label{fig:sim-continuous}
\end{figure}

\begin{table}[p]
\caption{Results of estimation metrics for coefficients using NeuralPLSI across different true link functions for binary outcome with $N=500$ and $N=2000$. SD = Standard Deviation, SE = Standard Error, and CP = Coverage Probability.}
\label{sim:results-bin-combined}
\centering
\footnotesize
\begin{tabular}{ccrrrrrrrr}
\toprule
\multirow{2}{*}{\textbf{$g$-function}} & \multirow{2}{*}{\textbf{Parameter}} &
\multicolumn{4}{c@{\hskip 12pt}}{\textbf{NeuralPLSI ($N=500$)}} &
\multicolumn{4}{c}{\textbf{NeuralPLSI ($N=2000$)}} \\
\cmidrule(lr){3-6} \cmidrule(lr){7-10}
& & \textbf{Bias} & \textbf{SD} & \textbf{SE} & \textbf{CP} & \textbf{Bias} & \textbf{SD} & \textbf{SE} & \textbf{CP} \\
\midrule
\multirow{11}{*}{Linear} 
& $\beta_1$  & -0.0934 & 0.1643 & 0.1514 & 0.9400 & -0.0227 & 0.0506 & 0.0562 & 0.9800 \\
& $\beta_2$  & -0.0693 & 0.1555 & 0.1820 & 1.0000 & 0.0039 & 0.0624 & 0.0661 & 1.0000 \\
& $\beta_3$  & 0.1231 & 0.2139 & 0.1761 & 0.9600 & 0.0030 & 0.0702 & 0.0641 & 0.9800 \\
& $\beta_4$  & -0.0408 & 0.2407 & 0.1829 & 1.0000 & -0.0017 & 0.0824 & 0.0701 & 0.9600 \\
& $\beta_5$  & 0.0147 & 0.1724 & 0.1743 & 1.0000 & -0.0137 & 0.0692 & 0.0725 & 0.9800 \\
& $\beta_6$  & 0.0253 & 0.1667 & 0.1721 & 1.0000 & 0.0032 & 0.0618 & 0.0726 & 1.0000 \\
& $\beta_7$  & 0.0012 & 0.1425 & 0.1742 & 1.0000 & -0.0012 & 0.0595 & 0.0733 & 1.0000 \\
& $\beta_8$  & 0.0068 & 0.2208 & 0.1736 & 1.0000 & 0.0041 & 0.0814 & 0.0720 & 0.9200 \\
& $\gamma_1$  & -0.3248 & 0.1741 & 0.1924 & 0.6000 & -0.0449 & 0.0689 & 0.0780 & 0.9800 \\
& $\gamma_2$  & 0.1407 & 0.1270 & 0.1337 & 0.9400 & 0.0077 & 0.0661 & 0.0635 & 0.9600 \\
& $\gamma_3$  & -0.1240 & 0.1475 & 0.1337 & 0.9000 & -0.0123 & 0.0704 & 0.0622 & 1.0000 \\
\cdashline{1-10}
\multirow{11}{*}{s-Shaped} 
& $\beta_1$  & -0.0156 & 0.1037 & 0.0749 & 0.9600 & -0.0066 & 0.0184 & 0.0253 & 0.9800 \\
& $\beta_2$  & -0.0265 & 0.0627 & 0.0902 & 1.0000 & -0.0009 & 0.0221 & 0.0306 & 1.0000 \\
& $\beta_3$  & 0.0205 & 0.0703 & 0.0807 & 1.0000 & -0.0059 & 0.0278 & 0.0289 & 0.9800 \\
& $\beta_4$  & -0.0034 & 0.1377 & 0.0897 & 0.9800 & -0.0012 & 0.0332 & 0.0323 & 0.9600 \\
& $\beta_5$  & -0.0125 & 0.0928 & 0.0827 & 0.9600 & 0.0007 & 0.0306 & 0.0338 & 1.0000 \\
& $\beta_6$  & 0.0078 & 0.0746 & 0.0813 & 1.0000 & 0.0030 & 0.0340 & 0.0336 & 1.0000 \\
& $\beta_7$  & 0.0033 & 0.0728 & 0.0823 & 1.0000 & 0.0030 & 0.0338 & 0.0333 & 0.9800 \\
& $\beta_8$  & -0.0070 & 0.1078 & 0.0822 & 1.0000 & 0.0024 & 0.0317 & 0.0333 & 0.9800 \\
& $\gamma_1$  & -0.3545 & 0.1379 & 0.2049 & 0.5800 & -0.0789 & 0.0903 & 0.0931 & 0.9000 \\
& $\gamma_2$  & 0.1488 & 0.1142 & 0.1510 & 0.9600 & 0.0199 & 0.0755 & 0.0775 & 1.0000 \\
& $\gamma_3$  & -0.1585 & 0.1170 & 0.1490 & 0.9000 & -0.0146 & 0.0838 & 0.0768 & 0.9800 \\
\cdashline{1-10}
\multirow{11}{*}{Sigmoid} 
& $\beta_1$  & -0.0114 & 0.0690 & 0.0981 & 0.9800 & -0.0071 & 0.0281 & 0.0332 & 0.9800 \\
& $\beta_2$  & -0.0449 & 0.0885 & 0.1163 & 1.0000 & 0.0052 & 0.0315 & 0.0396 & 1.0000 \\
& $\beta_3$  & 0.0518 & 0.0999 & 0.1082 & 1.0000 & 0.0012 & 0.0396 & 0.0376 & 0.9600 \\
& $\beta_4$  & 0.0278 & 0.1119 & 0.1160 & 1.0000 & -0.0093 & 0.0462 & 0.0427 & 0.9600 \\
& $\beta_5$  & 0.0015 & 0.0963 & 0.1097 & 0.9800 & 0.0022 & 0.0332 & 0.0439 & 1.0000 \\
& $\beta_6$  & -0.0109 & 0.0906 & 0.1081 & 0.9800 & 0.0053 & 0.0417 & 0.0436 & 0.9800 \\
& $\beta_7$  & 0.0165 & 0.0902 & 0.1096 & 0.9800 & 0.0014 & 0.0376 & 0.0435 & 0.9800 \\
& $\beta_8$  & -0.0160 & 0.0899 & 0.1102 & 0.9800 & -0.0025 & 0.0475 & 0.0428 & 0.9600 \\
& $\gamma_1$  & -0.3278 & 0.0846 & 0.1974 & 0.5800 & -0.0696 & 0.0747 & 0.0838 & 0.9200 \\
& $\gamma_2$  & 0.1260 & 0.0915 & 0.1433 & 0.9200 & 0.0142 & 0.0662 & 0.0693 & 0.9600 \\
& $\gamma_3$  & -0.1287 & 0.1030 & 0.1397 & 0.8800 & -0.0078 & 0.0736 & 0.0686 & 0.9800 \\
\bottomrule
\end{tabular}
\end{table}

\begin{figure}[p]
    \centering
    \includegraphics[width=0.95\linewidth]{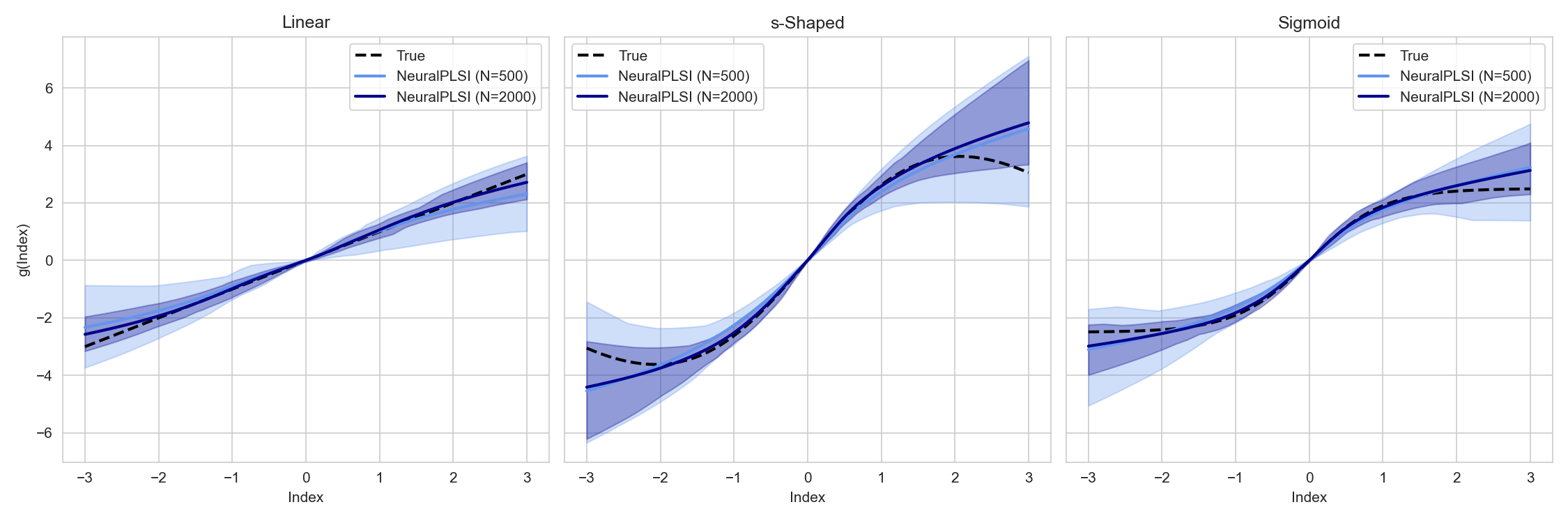}
    \caption{Estimated $g$ functions for the simulation study. }
    \label{fig:sim-binary}
\end{figure}

\begin{table}[p]
\caption{Results of estimation metrics for coefficients using NeuralPLSI across different true link functions for survival outcome with $N=500$ and $N=2000$. SD = Standard Deviation, SE = Standard Error, and CP = Coverage Probability.}
\label{sim:results-cox-combined}
\centering
\footnotesize
\begin{tabular}{ccrrrrrrrr}
\toprule
\multirow{2}{*}{\textbf{$g$-function}} & \multirow{2}{*}{\textbf{Parameter}} &
\multicolumn{4}{c@{\hskip 12pt}}{\textbf{NeuralPLSI ($N=500$)}} &
\multicolumn{4}{c}{\textbf{NeuralPLSI ($N=2000$)}} \\
\cmidrule(lr){3-6} \cmidrule(lr){7-10}
& & \textbf{Bias} & \textbf{SD} & \textbf{SE} & \textbf{CP} & \textbf{Bias} & \textbf{SD} & \textbf{SE} & \textbf{CP} \\
\midrule
\multirow{11}{*}{Linear} 
& $\beta_1$  & -0.0264 & 0.0728 & 0.1217 & 1.0000 & -0.0060 & 0.0337 & 0.0322 & 0.9400 \\
& $\beta_2$  & -0.0149 & 0.0818 & 0.1457 & 1.0000 & 0.0022 & 0.0408 & 0.0381 & 0.9400 \\
& $\beta_3$  & 0.0403 & 0.1137 & 0.1369 & 0.9200 & -0.0007 & 0.0425 & 0.0360 & 0.9600 \\
& $\beta_4$  & 0.0194 & 0.1024 & 0.1435 & 1.0000 & -0.0074 & 0.0416 & 0.0401 & 0.9800 \\
& $\beta_5$  & -0.0359 & 0.0864 & 0.1316 & 1.0000 & -0.0054 & 0.0383 & 0.0421 & 1.0000 \\
& $\beta_6$  & -0.0086 & 0.1034 & 0.1296 & 1.0000 & 0.0022 & 0.0416 & 0.0409 & 1.0000 \\
& $\beta_7$  & 0.0022 & 0.0992 & 0.1312 & 0.9800 & 0.0031 & 0.0441 & 0.0417 & 0.9400 \\
& $\beta_8$  & -0.0005 & 0.1142 & 0.1305 & 1.0000 & -0.0007 & 0.0457 & 0.0419 & 0.9400 \\
& $\gamma_1$  & -0.2094 & 0.0837 & 0.1695 & 0.7200 & -0.0437 & 0.0433 & 0.0472 & 0.9400 \\
& $\gamma_2$  & 0.0810 & 0.0772 & 0.0997 & 0.8600 & 0.0131 & 0.0406 & 0.0380 & 1.0000 \\
& $\gamma_3$  & -0.0775 & 0.0793 & 0.0992 & 0.8600 & -0.0104 & 0.0332 & 0.0382 & 0.9800 \\
\cdashline{1-10}
\multirow{11}{*}{s-Shaped} 
& $\beta_1$  & -0.0258 & 0.1222 & 0.0643 & 0.9800 & -0.0013 & 0.0143 & 0.0144 & 0.9600 \\
& $\beta_2$  & -0.0308 & 0.1474 & 0.0842 & 0.9800 & -0.0002 & 0.0160 & 0.0175 & 0.9800 \\
& $\beta_3$  & 0.0382 & 0.1812 & 0.0717 & 1.0000 & 0.0004 & 0.0200 & 0.0164 & 0.9600 \\
& $\beta_4$  & 0.0072 & 0.0583 & 0.0760 & 1.0000 & -0.0014 & 0.0177 & 0.0184 & 0.9600 \\
& $\beta_5$  & -0.0075 & 0.0407 & 0.0667 & 0.9800 & 0.0013 & 0.0185 & 0.0190 & 1.0000 \\
& $\beta_6$  & -0.0059 & 0.0450 & 0.0594 & 1.0000 & -0.0005 & 0.0209 & 0.0185 & 0.9600 \\
& $\beta_7$  & 0.0025 & 0.0444 & 0.0633 & 1.0000 & 0.0019 & 0.0177 & 0.0189 & 0.9600 \\
& $\beta_8$  & -0.0166 & 0.0949 & 0.0621 & 1.0000 & -0.0001 & 0.0205 & 0.0189 & 0.9400 \\
& $\gamma_1$  & -0.2446 & 0.1678 & 0.1636 & 0.6200 & -0.0543 & 0.0432 & 0.0491 & 0.9400 \\
& $\gamma_2$  & 0.1108 & 0.1208 & 0.1001 & 0.8400 & 0.0194 & 0.0355 & 0.0396 & 0.9800 \\
& $\gamma_3$  & -0.1074 & 0.1214 & 0.0991 & 0.8200 & -0.0143 & 0.0325 & 0.0393 & 0.9800 \\
\cdashline{1-10}
\multirow{11}{*}{Sigmoid} 
& $\beta_1$  & -0.0297 & 0.1135 & 0.0758 & 0.9800 & -0.0025 & 0.0188 & 0.0196 & 0.9400 \\
& $\beta_2$  & -0.0166 & 0.1020 & 0.0951 & 0.9600 & -0.0006 & 0.0230 & 0.0238 & 0.9600 \\
& $\beta_3$  & 0.0253 & 0.1336 & 0.0849 & 1.0000 & 0.0011 & 0.0263 & 0.0222 & 0.9600 \\
& $\beta_4$  & -0.0066 & 0.1235 & 0.0896 & 1.0000 & -0.0009 & 0.0254 & 0.0250 & 0.9400 \\
& $\beta_5$  & -0.0151 & 0.0718 & 0.0801 & 0.9800 & 0.0013 & 0.0250 & 0.0261 & 1.0000 \\
& $\beta_6$  & -0.0032 & 0.0788 & 0.0755 & 1.0000 & 0.0001 & 0.0285 & 0.0252 & 0.9600 \\
& $\beta_7$  & 0.0046 & 0.0669 & 0.0779 & 0.9800 & 0.0022 & 0.0260 & 0.0257 & 0.9400 \\
& $\beta_8$  & 0.0005 & 0.1118 & 0.0770 & 1.0000 & -0.0005 & 0.0273 & 0.0259 & 0.9400 \\
& $\gamma_1$  & -0.2370 & 0.1580 & 0.1637 & 0.7200 & -0.0485 & 0.0423 & 0.0479 & 0.9400 \\
& $\gamma_2$  & 0.1045 & 0.1132 & 0.0963 & 0.9000 & 0.0151 & 0.0403 & 0.0386 & 1.0000 \\
& $\gamma_3$  & -0.1035 & 0.1121 & 0.0972 & 0.8600 & -0.0100 & 0.0310 & 0.0387 & 0.9800 \\
\bottomrule
\end{tabular}
\end{table}

\begin{figure}[p]
    \centering
    \includegraphics[width=0.95\linewidth]{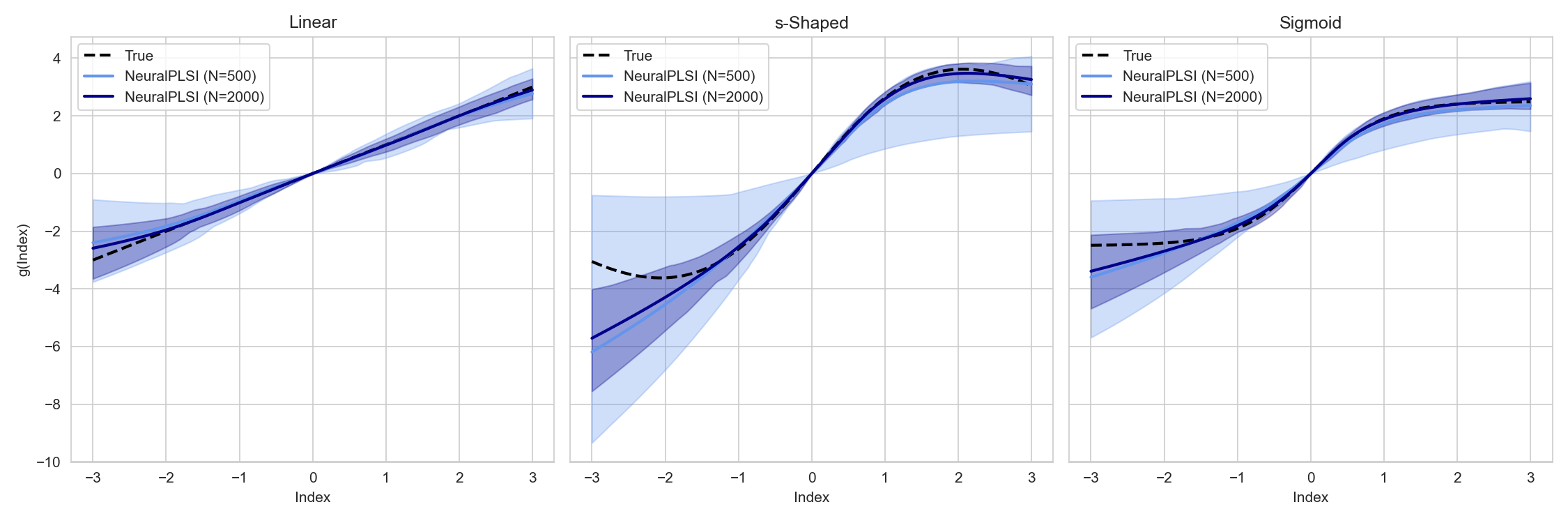}
    \caption{Estimated $g$ functions for the simulation study. }
    \label{fig:sim-cox}
\end{figure}

\end{document}